\documentclass[conference]{IEEEtran}
\usepackage[applemac]{inputenc}
\usepackage[english,american]{babel}
\usepackage{amsmath}
\usepackage{amssymb}
\usepackage{amsthm}
\usepackage{thmtools}
\usepackage{balance}
\usepackage{siunitx}
\usepackage{cite} % Interval of citations
\usepackage{hyperref} % Clickable references
\usepackage{tikz}
\usepackage{pgfplots}
\usepackage{graphicx}
\ifCLASSOPTIONcompsoc
\usepackage[caption=false,font=normalsize,labelfont=sf,textfont=sf]{subfig}
\else
\usepackage[caption=false,font=footnotesize]{subfig}
\fi

\usepackage{ctable}

\newtheoremstyle{mydef}
{3pt}		% ?Space above?
{3pt}		% ?Space below?
{}		% ?Body font?
{}		% ?Indent amount?1
{\itshape}	% ?Theorem head font?
{:}		% ?Punctuation after theorem head?
{.5em}	% ?Space after theorem head?2
{}		% ?Theorem head spec (can be left empty, meaning ?normal?)?

\theoremstyle{mydef}

\newcommand{\otoprule}{\midrule[\heavyrulewidth]}
\newcommand{\IE}{i.e., } % id est				
\newcommand{\figref}[1]{Fig.\,\ref{fig:#1}}
\newcommand{\RN}[1]{%
	\textup{\uppercase\expandafter{\romannumeral#1}}%	 Roman uppercase number
}
\newcommand{\Rho}{\operatorname{P}} % capital rho
\newcommand{\Po}[1]{\ensuremath{\text{\textsf{Po}}(#1)}} % Poisson distribution
\newcommand{\N}{n} 				% n the number of slots in a frame	
\newcommand{\NRX}{\N_{\text{RX}}} % n_rx the number of slots in receiver memory
\newcommand{\T}{i} 				 % index i (time)	
\newcommand{\G}{g}				% system load g	

\newcommand{\Graph}{\mathcal{G}}% graph G 
\newcommand{\EdgeG}{\mathcal{E}}% edge E
\newcommand{\Variables}{\mathcal{V}}% VN V
\newcommand{\Checks}{\mathcal{C}} % CN C
\newcommand{\M}{M}				% active users
\newcommand{\Mtime}{M_{\T}} 	% active users at time \T
\newcommand{\pii}{p_{i\rightarrow i}}
\newcommand{\pij}{p_{i\rightarrow j}}
\newcommand{\qii}{q_{i\rightarrow i}}
\newcommand{\qij}{q_{i\rightarrow j}}
\newcommand{\qji}{q_{j \rightarrow i}}
\newcommand{\pji}{p_{j \rightarrow i}}

\newcommand{\qavi}{\tilde{q}_i}
\newcommand{\pavi}{\tilde{p}_i}
\newcommand{\Jset}{\mathcal{J}_i}
\newcommand{\Kset}{\mathcal{K}_i}
\newcommand{\CNddother}{\Rho^{i\rightarrow \Kset}}
\newcommand{\CNddotherr}{\Rho^{i\rightarrow \Kset}_{r_2}}
\newcommand{\CNedotherr}{\rho^{i\rightarrow \Kset}_{r_2}}
\newcommand{\CNddsame}{\Rho^{i\rightarrow i}}
\newcommand{\CNddsamer}{\Rho^{i\rightarrow i}_{r_1}}
\newcommand{\CNedsamer}{\rho^{i\rightarrow i}_{r_1}}

\newcommand{\CNdd}{\Rho }
\newcommand{\CNed}{\rho }
\newcommand{\VNdd}{\Lambda }
\newcommand{\VNed}{\lambda }
\newcommand{\VNedTilde}{\lambda^{i\rightarrow \Jset}}
\newcommand{\VSameTilde}{\lambda^{i\rightarrow i}}
\newcommand{\VNddOpt}{\Lambda^{\star}}
\newcommand {\define}{\stackrel{\Delta}{=}}
\newcommand{\PLR}{\bar{p}} % PLR

\newcommand{\user}{u}	% User u	
\newcommand{\Sset}{\mathcal{S}} % Stopping Set S
\newcommand{\numCNs}{\mu \left(\Sset\right)} % number of CNs in a stopping set
\newcommand{\numVNs}{\nu \left(\Sset\right)} % number of VNs in a stopping set
\newcommand{\numVNsL}{v_l\left(\Sset\right)} % number of degree-l VNs in a stopping set
\newcommand{\numVNsD}{v_d\left(\Sset\right)} % number of degree-l VNs in a stopping set
\newcommand{\iso}{c(\Sset)} % number of isomorphisms of a Stopping set
\newcommand{\A}{\mathcal{A}} % The set of all stopping Sets A
\newcommand{\Astar}{\mathcal{A}^{\star}} % The set of all stopping Sets A
\newcommand{\ps}{\Pr\left(\user \in \Sset |m\right)} %
\newcommand{\aS}{a(\Sset,m)}
\newcommand{\bS}{b(\Sset)}
\newcommand{\bSFS}{b_{\mathrm{FS}}(\Sset)}
\newcommand{\dS}{d(\Sset)}
\newcommand{\dSFS}{d_{\mathrm{FS}}(\Sset)}

\newcommand{\gstar}{\G^{\star}}
\newcommand{\gstarfaBE}{\G^{\star}_{\text{FA,B}}}
\newcommand{\gstarfanoBE}{\G^{\star}_{\text{FA}}}
\newcommand{\gstarfs}{\G^{\star}_{\text{FS}}}

\IEEEoverridecommandlockouts

\title{Asymptotic and Finite Frame Length Analysis of  Frame Asynchronous Coded Slotted ALOHA}
\begin{document}
\author{
\IEEEauthorblockN{Erik Sandgren, Alexandre Graell i Amat, and Fredrik Br\"annstr\"om}
%\thanks{This work was partially funded by the Swedish Research Council under grants 2011-5950 and 2011-5961.}
%\IEEEauthorblockA{Department of Signals and Systems, Chalmers University of Technology, Gotheburg, Sweden}\\Email: \textit{erik.sandgren92@gmail.com,  \{alexandre.graell, fredrik.brannstrom\}@chalmers.se}}
\IEEEauthorblockA{Department of Signals and Systems, Chalmers University of Technology, Gothenburg, Sweden}
}
\maketitle
%%%%%%%%%%%%%%%%%%%%%%%%%%%%%%%%%%%%%%%%%%%%%%%%% ABSTRACT %%%%%%%%%%%%%%%%%%%%%%%%%%%%%%%%%%%%%%%%%%%%%%%%%%%%%%%%%
\begin{abstract}
We consider a frame-asynchronous coded slotted ALOHA (FA-CSA) system where users join according to a Poisson random process. In contrast to standard frame-synchronous CSA (FS-CSA), 
%users transmit a first replica of their message in the slot following their activation 
when a user joins the system, it transmits a first replica of its message in the following slot and other replicas uniformly at random in a number of subsequent slots. We derive the (approximate) density evolution that characterizes the  asymptotic performance of FA-CSA when the frame length goes to infinity. We show that, if the receiver can monitor the system before users start transmitting, a boundary effect similar to that of spatially-coupled codes occurs, which greatly improves the decoding threshold as compared to FS-CSA. We also derive analytical approximations of the error floor (EF) in the finite frame length regime. We show that  FA-CSA yields in general lower EF, better performance in the waterfall region, and lower average delay, as compared to FS-CSA.
\end{abstract}

%%%%%%%%%%%%%%%%%%%%%%%%%%%%%%%%%%%%%%%%%%%%%% INTRODUCTION %%%%%%%%%%%%%%%%%%%%%%%%%%%%%%%%%%%%%%%%%%%%%%%%%%%%%%%%
\section{Introduction}
Recently, a new class of uncoordinated multiple access techniques, named  coded slotted ALOHA (CSA), has emerged and attracted much interest. CSA is based on the idea of classical slotted ALOHA. The key innovation of CSA is that users replicate each packet over several slots and decoding is performed over a sequence of slots using successive interference cancellation. This idea was first introduced in \cite{casini2007contention} and further developed in \cite{liva2011graph}, where the connection between CSA and codes on graphs was established.
%Many other works have considered extensions to CSA since then.
%This connection enables the use of already existing tools for the analysis of CSA.

A great advantage of CSA is that it can provide high throughput and reliability without the use of an automatic repeat request scheme \cite{ivanov2015error}. %Following the introduction of CSA most works on the topic have focused on the analysis of its asymptotic performance and on improving its throughput. 
However, depending on the user model of the CSA system, the delay may be relatively high\cite{meloni2012sliding}. To improve the delay performance, a frame-asynchronous CSA (FA-CSA) system was proposed in \cite{meloni2012sliding}. In FA-CSA,  when a user joins the system, it transmits a first replica of its message in the following slot. Remaining replicas are randomly distributed in a number of subsequent slots. This system is in contrast with classical CSA in which users are frame synchronous (FS) and communication takes place within a predetermined number of slots, called \textit{frame}. Simulation results in \cite{meloni2012sliding} show that, in addition to improve the average delay, FA-CSA also outperforms FS-CSA in terms of throughput. %Moreover, the effect of using a sliding-window decoder for FA-CSA was investigated in \cite{meloni2012sliding}.

In this paper, we analyze the asymptotic and the finite frame length performance of FA-CSA in terms of packet loss rate (PLR).  We derive the (approximate) density evolution (DE) equations that govern the asymptotic behavior of FA-CSA. We show that, if the receiver can monitor the system before users start transmitting, a boundary effect similar to that of spatially-coupled codes occurs. This effect greatly improves the decoding threshold as compared to FS-CSA.  Furthermore, we derive analytical approximations of the PLR in the finite frame length regime, based on the framework introduced in \cite{ivanov2015error} and \cite{ivanov2015broadcast}, in order to predict the error floor (EF) of FA-CSA. It is shown that, in general, FA-CSA achieves superior performance in both the EF and waterfall (WF) regions compared to FS-CSA. In addition, we show that FA-CSA achieves a lower average delay.
% We show that the EF of FA-CSA is in general lower than that of FS-CSA. Finally we present a comparison of FA-CSA and FS-CSA, based on our analysis and simulation results. 

%%%%%%%%%%%%%%%%%%%%%%%%%%%%%%%%%%%%%%%%% SYSTEM MODEL %%%%%%%%%%%%%%%%%%%%%%%%%%%%%%%%%%%%%%%%%%%%%%%%%%%%%%%%%%%%%
\section{System Model}
We consider a CSA system where users are slot synchronized and transmit to a common receiver.  A user that joins the system selects a repetition factor $l$ randomly according to a predefined degree distribution \cite{liva2011graph}. The user then maps its message into a physical layer packet and transmits $l$ copies (called replicas) of the packet in randomly chosen slots. We say that a user that repeats its packet $l$ times is a degree-$l$ user and that a slot in which $r$ packets collide is a degree-$r$ slot.
%A user that becomes active picks a repetition degree $l$ from a predefined degree distribution. The $l$ message-replicas will be placed in a set of $\N$ slots of equal duration. Where $\N$ is referred to as the frame length. Every sent packet-replica is received by a common receiver.
\label{sec:SystemModel}
We assume that users join the system on a slot basis according to a Poisson process. Let $K$ denote the number of users that join in a slot. Then $K$ is a Poisson distributed random variable (RV) with mean $\G$ users per slot, $K \sim \Po{\G}$.
% where  $\G$ is the average system load in users per slot
 The probability that $k$ users join in a given slot is thus,
\begin{equation}
\Pr(K=k) = \frac{\G^k}{k!}e^{-\G}.
\end{equation}
\subsection{Frame Synchronous Coded Slotted ALOHA}
In FS-CSA, communication takes place during a frame consisting of $\N$ slots. A user that joins the system waits until the next frame, where it becomes active and transmits its $l$ replicas in randomly chosen slots of that frame. We denote by $\M \sim \Po{\N\G}$ the RV representing the number of active users per frame. 

Decoding is performed by the receiver on a slot-by-slot basis. Assume the decoding of slot $i$. First, the interference caused by packets for which replicas in previous slots have already been decoded is canceled from the slot. This is possible because every packet contains pointers to all its replicas. The receiver then checks if slot $i$ is a degree-one slot, and if not, the decoding of slot $i$ is stopped. Otherwise, the packet in slot $i$ is decoded and the interference from all its replicas canceled from the corresponding past slots.  The receiver then proceeds to iteratively find any degree-one slots in its memory, decode the packets in these slots, and cancel the interference of all replicas of the decoded packets. This process continues until no new degree-one slots appear or a maximum number of iterations is reached. 

%Decoding is performed by the receiver on a slot basis. First, the receiver cancels the interference of any previously decoded replicas from the current slot. It then checks if the current slot is degree-one, if not, decoding is stopped. Otherwise, the packet in the current slot is decoded and the interference from all its replicas canceled from their corresponding slots (either past or future). This is possible because every packet contains pointers to all its replicas.  The receiver then proceeds to iteratively find any degree-one slots in its memory (and in the current frame), decode the packets in degree-one slots, and cancel the interference of all replicas of the decoded packets. This process continues until no new degree-one slots appear or for a maximum number of iterations. 
 
Throughout the paper we assume perfect interference cancellation (IC), \IE all replicas of a decoded packet can be perfectly canceled from their slots. In \cite{casini2007contention,liva2011graph} actual (low complexity) IC was implemented with little performance degradation as compared to perfect IC.% for signal-to-noise ratios as low as $\SI{2}{\decibel}$.

\subsection{Frame Asynchronous Coded Slotted ALOHA}
\begin{figure}[]
	\centering
	\includegraphics[width=\columnwidth]{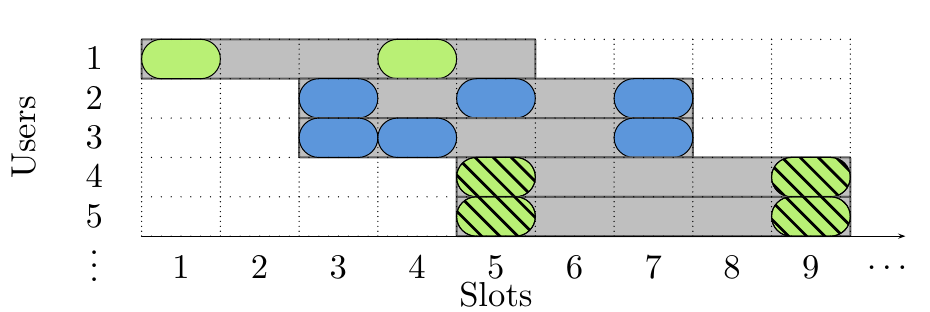}
	\vspace{-4ex}
	\caption{FA-CSA with boundary effect, $\N=5$, $\G=0.5$, and $\VNdd(x)=0.6x^2+0.4x^3$. Green boxes are replicas of degree-2 users and blue boxes are replicas of degree-3 users. The striped boxes constitute a stopping set.}
	\label{fig:examp}
	\vspace{-2ex}
\end{figure}
%slots are not arranged into common frames as in  FS-CSA
In FA-CSA, when a degree-$l$ user joins the system it transmits a first replica in the following slot. The remaining $l-1$ replicas are distributed uniformly within the $\N-1$ subsequent slots. Therefore, contrary to FS-CSA, slots are not arranged in common frames. We call the $\N$ slots in which a user can transmit its \textit{local frame} and say that a user is active the entire duration of its local frame. Decoding is performed in a similar manner as for FS-CSA, with the difference that the receiver does not only consider  slots in the current local frame, but the entire history of the system. In practice, the memory (in number of slots) of the receiver, $\NRX$, cannot be arbitrarily large. It is in general sufficient to set $\NRX=5\N$ without loss of performance \cite{meloni2012sliding}. 

Let $\Mtime$ denote the number of active users in the $i$th slot, $i \geq 1$. We consider two different models for the initialization of the system, \IE for  $1\leq \T \leq \N$. The first model assumes that there are no active users at $\T=0$. In this case
\begin{equation}
\Mtime \sim \begin{cases} 
\Po{\T\G} & \text{ for } 1 \leq \T <\N \\
\Po{\N\G} & \text{ for } \T \geq \N
\end{cases},
\label{eq:init1}
\end{equation}
and we say that a \textit{boundary effect} is present for this model. 

The second model assumes that there are already $\M \sim \Po{\N\G}$ active users at $\T=0$. Thus,
\begin{equation}
\Mtime \sim \Po{\N\G} \quad \text{ for all } \T \geq 1.
\label{eq:init2}
\end{equation}
%In contrast with the first model, this model does \textit{not} have a boundary effect.

An example of FA-CSA with boundary effect is depicted in \figref{examp}. In the example, users 4 and 5 cannot be resolved.

\begin{figure}[]
	\centering
	\includegraphics[width=0.7\columnwidth]{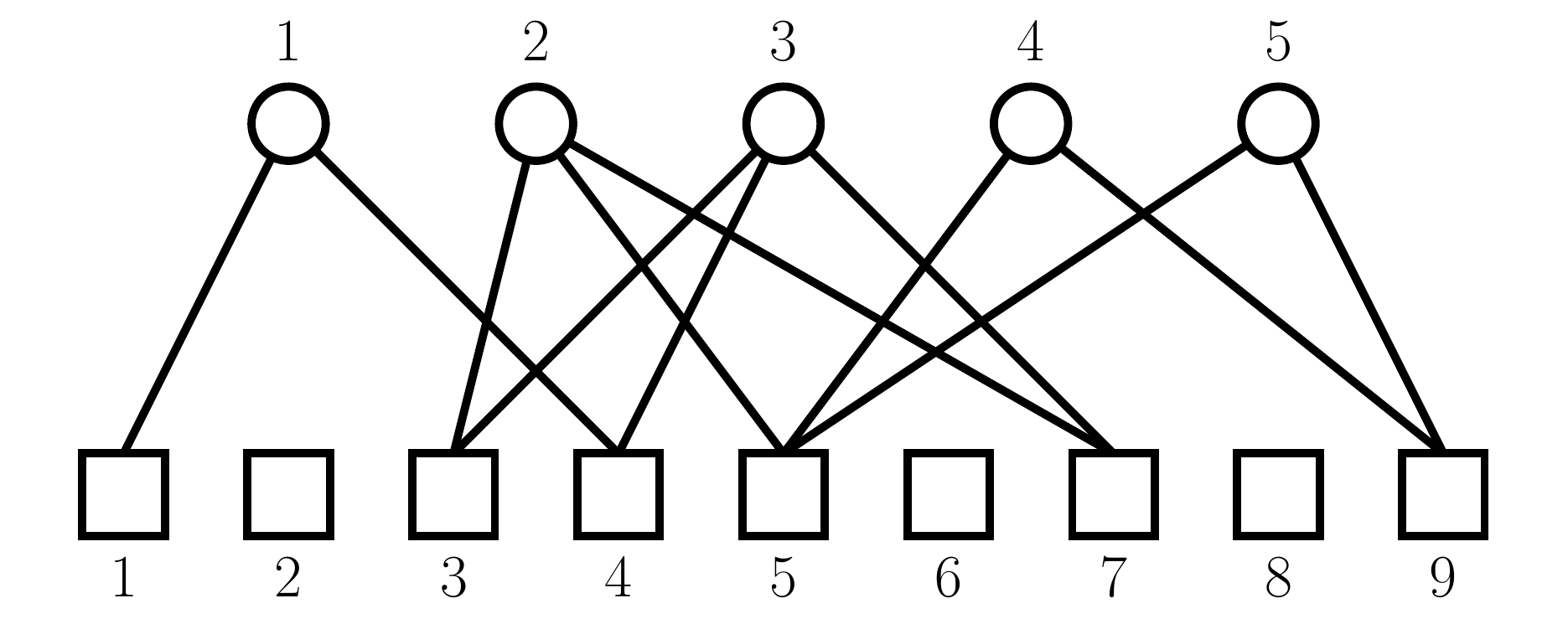}
	\vspace{-2ex}
	\caption{Equivalent graph representation of \figref{examp}, where VNs (circles) represent users and CNs (squares) represent slots.}
	\label{fig:graph_eqv}
	\vspace{-2ex}
\end{figure}

 CSA can be represented by means of a bipartite graph  $\Graph=\{\Variables,\Checks,\EdgeG\}$, where $\Variables$ is the set of variable nodes (VNs), $\Checks$ is the set of check nodes (CNs), and $\EdgeG$ is the set of edges connecting the VNs and CNs. VNs represent users and CNs represent slots. There is an edge $e_{i \rightarrow j} \in \EdgeG$ from VN $i$ to CN $j$ if user $i$ transmits a replica in slot $j$. Decoding of CSA can be viewed as message passing over the edges of the underlying graph \cite{liva2011graph}.  The degree of a node is equal to the number of edges incident to the node. In \figref{graph_eqv}, the  graph representation of the scenario depicted in \figref{examp} is shown.

%In the remainder of this paper  the terms  user (slot) and VN (CN) are used interchangeably

%%%%%%%%%%%%%%%%%%%%%%%%%%%%%%%%%%%%%%%%%%%%%%% ASYMPTOTIC ANALYSIS %%%%%%%%%%%%%%%%%%%%%%%%%%%%%%%%%%%%%%%%%%%%%%%%%%%%%%%
\section{Asymptotic Performance Analysis} 
\label{sec:DE}
In the asymptotic regime, \IE when $\N \rightarrow \infty$, CSA exhibits a threshold behavior: all users can be resolved if the system operates below a given system load, called threshold and denoted by $\gstar$. %However, if the system operates above the threshold, the PLR, defined as the average number of unresolvable users and denoted by $\PLR$,  will be bounded away from zero. 
The threshold can be obtained via DE. 	 

%\subsection{Degree Distributions}

We define  the node-perspective VN and CN degree distributions as
\begin{equation}
\label{eq:NPdegreedist}
\VNdd(x)= \sum_{l}^{} \VNdd_l x^l\quad \text{and } \quad  	\CNdd(x)=\sum_{r}^{} \CNdd_r x^r,
\end{equation}
respectively,
where $\VNdd_l$ is the probability that a VN has degree $l$  and $\CNdd_r$ is the probability that a CN has degree $r$. $\CNdd(x)$ is determined by $\VNdd(x)$ and the system model.  We will also consider the edge-perspective degree distributions
\begin{equation}
	\VNed(x)  = \sum_{l}^{}\VNed_l x^{l-1}\quad \text{and } \quad   \CNed(x) =\sum_{r}^{} \CNed_r x^{r-1},  %=\frac{\VNdd'(x)}{\VNdd'(1)}                   =\frac{\CNdd'(x)}{\CNdd'(1)}
\end{equation}
for VNs and CNs,
respectively,  where $\lambda(x)=\Lambda'(x)/\Lambda' (1)$, $\VNed_l = \VNdd_l l / \VNdd'(1)$, $\rho(x)=\CNdd'(x)/\CNdd '(1)$, and $\CNed_r = \CNdd_r r /  \CNdd'(1)$. Here $\VNed_l$ is the probability that an edge is connected to a degree-$l$ VN, $\CNed_r$ is the probability that an edge is connected to a degree-$r$ CN, and $f'$ is the derivative of the function $f$.

We assume an FA-CSA system with boundary effect. In this case, the first $\N$ CNs all have distinct degree distributions. This gives rise to different \textit{classes} of CNs and VNs. % that need to be treated separately in the DE. 
We call a class-$i$ CN a CN at position $i$. Similarly, a class-$i$ VN is a VN at position $i$. We also denote by $\pij$ the erasure probability from a class-$i$ VN to a class-$j$ CN, and by $\qij$ the erasure probability from a class-$i$ CN to a class-$j$ VN. 

\begin{figure}
	\centering
	\includegraphics[scale=0.4]{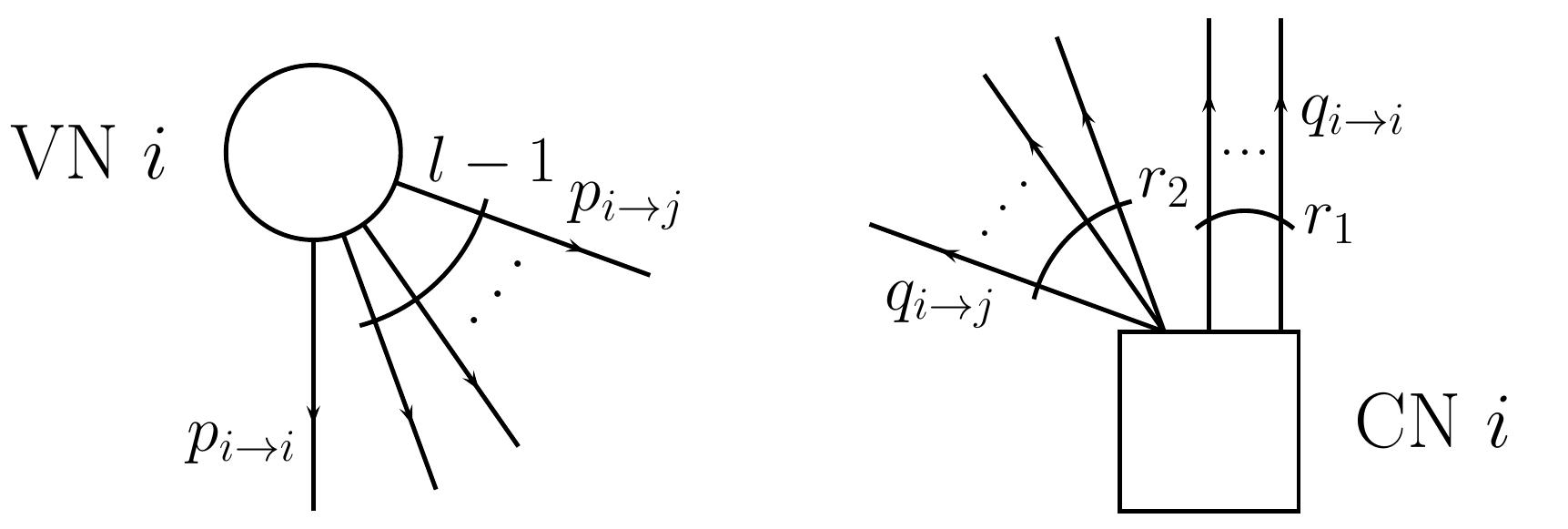}
	\vspace{-2ex}
	\caption{VN and CN of class $i$, and their corresponding connections.} % A class-$i$ VN has $1$ connection to a class-$i$ CN and $l-1$ connections to CNs of class $j > i$. A CN of class $i$ has $r_1$ connections to VNs of class $i$ and $r_2$ connections to VNs of type $j < i$.}
	\label{fig:DE_description}
	\vspace{-2ex}
\end{figure}

The graph connectivity of class-$i$ VNs and CNs is shown in \figref{DE_description}. A class-$i$ VN has $l-1$ connections to CNs at positions in the range $\Jset \define [i+1,i+n-1]$. Furthermore, it has one connection to a class-$i$ CN. Accordingly, we define the node-perspective degree distributions  $\Lambda^{i\rightarrow i}(x)=x$ and $\Lambda^{i\rightarrow \Jset}(x)=\sum_l\Lambda^{i\rightarrow \Jset}_lx^{l}=\sum_l\Lambda_lx^{l-1}$, where $\Lambda^{i\rightarrow \Jset}_l=\Lambda_{l+1}$ is the probability that a class-$i$ VN has $l$ connections to CNs at positions in the range $\Jset$. The corresponding edge-perspective degree distributions are $\VSameTilde(x)=1$ and 
$\VNedTilde(x)=(\Lambda^{i\rightarrow \Jset})'(x)/(\Lambda^{i\rightarrow \Jset})'(1)=\sum_l \lambda^{i\rightarrow \Jset}_lx^{l-2}$, with $\lambda^{i\rightarrow \Jset}_l= \VNdd_l(l-1)/\sum_{l}^{}\VNdd_l(l-1)$. On the other hand, a class-$i$ CN is connected to $r_1$ class-$i$ VNs and to $r_2$ VNs in the range $\Kset \define[\max(1,i-n+1),i-1]$. Correspondingly, we define the degree distributions $\CNddsame (x)=\sum_{r_1}\CNddsamer x^{r_1}$ and $\CNddother (x)=\sum_{r_2} \CNddotherr x^{r_2}$ as in (\ref{eq:NPdegreedist}), where $\CNddsamer$ is the probability that a class-$i$ CN has $r_1$ edges incident to class-$i$ VNs, and $\CNddotherr$ is the probability that a class-$i$ CN has $r_2$ connections to VNs in the range $\Kset$. $\CNddsame(x)$ simply follows the user model of the system, \IE it is a  Poisson distribution with mean $\G$. $\CNddother(x)$ is given by the sum of a number of Poisson distributions, each with mean $\frac{(\VNdd'(1)-1)\G}{\N-1}$. In all,
 \begin{equation}
\CNddsamer = \frac{e^{-\G}\G^{r_1}}{r_1!}  \quad \text{and} \quad  \CNddotherr = \frac{e^{-\mu_i}\mu_i^{r_2}}{r_2!},
\label{eq:cndd}
\end{equation}
where  
\begin{equation}
\mu_i  = \begin{cases} (\VNdd'(1)-1)\G \frac{i-1}{\N-1}& \text{ for } i < \N \\
(\VNdd'(1)-1)\G & \text{ for } i \geq \N   .
\end{cases}
\label{eq:mu}			
\end{equation}
The corresponding edge-perspective degree distributions are $\rho^{i\rightarrow i}(x)=\sum_{r_1}\CNedsamer x^{r_1-1}$ and $\rho^{i\rightarrow \Kset}(x)=\sum_{r_2}\CNedotherr x^{r_2-1}$, with
\begin{equation}
\CNedsamer = \frac{e^{-\G}\G^{r_1-1}}{(r_1-1)!}  \quad \text{and} \quad  \CNedotherr = \frac{e^{-\mu_i}\mu_i^{r_2-1}}{(r_2-1)!} .
\end{equation}

\subsection{Density Evolution}
We now derive the DE equations for FA-CSA. Note that since a class-$i$ VN is always connected to a class-$i$ CN, we must differentiate between \textit{edge types}, and thus update $\pii$, $\pij$, $\qii$, and $\qij$ in the DE separately. A message from a VN is in erasure if all incoming messages are in erasure, i.e.,
% outVN-edge is resolved if any of the VNs other $l-1$ edges were resolved in the previous iteration, that is
\begin{align}
\label{eq:pii}
\pii &= \sum_{l}^{}\VNdd_l \qavi^{\,l-1} = \Lambda^{i\rightarrow \Jset}(\qavi),\\%\frac{\VNdd \left( \qavi \right)}{\qavi},\\
\pij &=  \qii\sum_{l}^{} \VNedTilde_l \qavi^{\, l-2} =  \qii \VNedTilde \left( \qavi \right),%\frac{\qii}{\qavi } \VNedTilde \left( \qavi \right)
\label{eq:pij}
\end{align}
where  
%\begin{equation}
$\qavi = \frac{1}{\N - 1}\sum_{j \in \Jset}^{} \qji$. %\quad \VNedTilde_l = \frac{\VNdd_l(l-1)}{\sum_{l}^{}\VNdd_l(l-1)},
%\end{equation}

A message from a CN is not in erasure if none of the incoming $r_1+r_2-1$ messages is in erasure. Therefore,
\begin{IEEEeqnarray*}{rCl}
\qii &= 1 -  &\left(\sum_{r_1 = 1}^{\infty}\CNedsamer(1-\pii)^{r_1-1} \right)\times  \\
	&  &\left( \sum_{r_2=0}^{\infty} \CNddotherr (1-\pavi{})^{r_2} \right) \\
	&\stackrel{(a)}{=} 1- &\left(\sum_{r_1 = 0}^{\infty}\frac{e^{-\G}\G^{r_1}}{r_1!}(1-\pii)^{r_1} \right) \times \\
	& &\left( \sum_{r_2=0}^{\infty} \frac{e^{-\mu_i}\mu_i^{r_2}}{r_2!}(1-\pavi{})^{r_2} \right) \\
	 &= 1-  &e^{-\G\pii}e^{-\mu_i\pavi{}} \IEEEyesnumber,
	\label{eq:qii}
\end{IEEEeqnarray*}
where in $(a)$ we used $r'_1=r_1-1$ and $r'_1\leftarrow r_1$, and
\begin{equation}
\pavi = \frac{1}{k}\sum_{j \in \Kset}^{} \pji, \quad k= \begin{cases}
1 &  \text{if } i=1 \\
i-1 &  \text{if } 1<i<\N \\
\N - 1&  \text{if } i\geq \N
\end{cases}.
\end{equation} 
Similarly, 
\begin{IEEEeqnarray*}{rCl}
\qij&= 1 - &\left(\sum_{r_1 = 0}^{\infty}\CNddsamer(1-\pii)^{r_1} \right)\times\\
&  &\left( \sum_{r_2=1}^{\infty} \CNedotherr (1-\pavi{})^{r_2-1} \right) \\
&= 1-  &e^{-\G\pii}e^{-\mu_i\pavi{}}. \IEEEyesnumber
\label{eq:qij}
\end{IEEEeqnarray*}
Note that $\qii=\qij$, which follows from the properties of the Poisson distribution. For a general user model, however, $\qii\neq\qij$.

DE is now performed by iteratively updating \eqref{eq:pii}-\eqref{eq:pij} and \eqref{eq:qii}-\eqref{eq:qij}, with $\pii,\,\pij,\,\qii, \text{ and } \qij$  initialized to $1$. The PLR at position $i$ can be computed as $\PLR_i=\VNdd(\qavi)\qii / \qavi$. The threshold $\gstar$ is found by searching for the largest value of $g$ for which $\PLR_i$ converges to 0 for all positions. 

We remark that exact DE requires $\N \rightarrow \infty$. This would require to keep track of an infinite number of node classes, which is unfeasible in practice. Therefore, the thresholds computed in Section~\ref{sec:Results} must be seen as \emph{approximate} DE thresholds. However, we have found that it is sufficient to set $\N \approx 100$ and run DE over a chain of $20\N$ positions in order to obtain $\gstar$ with good precision. Considering larger values of $n$ does not change the obtained thresholds.

The DE equations for FA-CSA without boundary effect are identical to those of FA-CSA with boundary effect. However, \eqref{eq:pii}-\eqref{eq:qij} are iteratively updated only for $i>\N$. Note that in this case all CNs have identical degree distribution.% still $\pii$, $\pij$, $\qii$ and $\qij$ are initialized to 1 for all $i$. 

%For FS-CSA DE can be performed by simply tracking two parameters $p$ and $q$ \cite{liva2011graph}. $p$  is the average probability of an unresolved VN edge, and $q$ is the average probability of an unresolved CN-edge, 
%\begin{equation}
%		q=1-e^{-gp}, \quad \quad p = \VNed(q).
%\end{equation}
%where $\PLR=\VNdd(q)$.

%%%%%%%%%%%%%%%%%%%%%%%%%%%%%%%%%%%%%%%%%%%%% FINITE ANALYSIS %%%%%%%%%%%%%%%%%%%%%%%%%%%%%%%%%%%%%%%%%%%%%%%%%%%%%%%%%
\section{Finite Frame Length Analysis}
\label{sec:EF}
%In CSA the PLR depends predominantly on three parameters, the frame length $\N$, the system load $\G$ and the  degree distribution $\VNdd(x)$. As illustrated in Fig \ref{fig:examp},
Packet losses in CSA are caused by stopping sets \cite{ivanov2015error}. In this section, we analyze the PLR of FA-CSA by finding the probability of occurrence of stopping sets. We use the framework in \cite{ivanov2015broadcast} and \cite{ivanov2015error}, and extend it to FA-CSA. 
%In \cite{ivanov2015error} it was  shown that the EF for CSA-systems is dominated by \textit{minimal stopping sets}. 
For the analysis, we consider an FA-CSA system without boundary effect, since the boundary will have negligible impact on the EF.
%when running a finite frame length system for a long time
For a stopping set $\Sset$, let $\numCNs$ denote the number of CNs, $\numVNs$  the number of VNs, and $\numVNsL$  the number of degree-$l$ VNs. Moreover, we denote by $\iso$ the number of graph isomorphisms of $\Sset$ \cite[p.4]{bondy1976graph}. Unfortunately, there is no straightforward analytical expression for $\iso$. However, $\iso$ is tabulated in \cite[Table \RN{1}]{ivanov2015broadcast} along with $\numVNs$, $\numCNs$, and $\numVNsL$ for all 31 minimal stopping sets\footnote{A minimal stopping set is a stopping set that does not contain a nonempty stopping set of smaller size.} with $\numCNs\le 4$.
%user active at slot $i$

Let $\user$ represent a VN in the range $[i,i+n-1]$. Furthermore, let $\A$ denote the set of all stopping sets and $\Astar \subset \A$ a finite set of minimal stopping sets. An approximation of the PLR, $\PLR$, can with some abuse of notation be written as
\begin{IEEEeqnarray*}{rl}
\PLR & = \Pr\left(\bigcup_{\Sset \in \A} \user \in \Sset \right)\stackrel{(a)}{\leq} \sum_{\Sset \in \A}\Pr\left(\user \in \Sset \right) 
\stackrel{(b)}{\approx}\sum_{\Sset \in \Astar}\Pr\left(\user \in \Sset \right) \\
	& =\sum_{\Sset \in \Astar}\sum_{m=0}^{\infty}\frac{e^{-\N\G}(\N\G)^m}{m!}\ps.\IEEEyesnumber  \label{eq:PLR}
\end{IEEEeqnarray*}
In $(a)$ the probability is upper bounded using the union bound. In $(b)$ we consider a summation over a subset of stopping sets, $\Astar$, turning the upper bound into an approximation. We also take the expectation with respect to the RV that represents the number of active users $\M \sim \Po{\N\G}$ in slot $i+n-1$. 
%In the final step we introduce $\ps$, representing the probability that a user $\user$ is in the stopping set $\Sset$ when there are $m$ active users. We also take the expectation of $\ps$ with respect to the RV that represents the number of active users $\M \sim \Po{\N\G}$. In the following, we derive  $\ps$ with some abuse of notation. 

We now derive an approximation of $\ps$. In the sequel, with some abuse of language we use the terms slot and user instead of CN and VN, respectively. To simplify the derivation, we make the assumption that $\Sset$ spans at most $n$ slots. Without loss of generality, we consider the range $[i,i+n-1]$. We write $\ps$ as
% We assume that all users of a potential stopping set $\Sset$  are selected from the set of $m$ active users at slot $i$ and write,
%We now derive $\ps$ with some abuse of language \IE we use the terms slot and user instead of CN and VN respectively. To start with we express $\ps$ as, 
\iffalse
\begin{IEEEeqnarray*}{rCl}
\ps%&= \frac{\aS\bS\iso \fS}{\dS \fS} \cdot \frac{\numVNs}{m} \\
&= \frac{\aS\bS \iso}{\dS} \cdot \frac{\numVNs}{m},
\IEEEyesnumber
\label{eq:ps}
\end{IEEEeqnarray*}
\fi
\begin{IEEEeqnarray*}{rCl}
	\ps%&= \frac{\aS\bS\iso \fS}{\dS \fS} \cdot \frac{\numVNs}{m} \\
	&\approx \frac{\aS\bS \iso}{\dS} \cdot \frac{\numVNs}{m},
	\IEEEyesnumber
	\label{eq:ps}
\end{IEEEeqnarray*}
where $\aS$ is the expected number of ways to select $\numVNs$ users of the same degrees as the users in $\Sset$ from a set of $m$ users, $\bS$ is the number of ways to select the slots of $\Sset$ such that $u$ is in $\Sset$, 
%$\fS$ is the number of ways in which $m-\numVNs$ users, not in $\Sset$, can select slots for transmission, 
and $\dS$ is the total number of ways in which $\numVNs$ users (including $u$) of the same degrees as the users in $\Sset$ can select slots for their replicas. The fraction $\frac{\numVNs}{m}$ represents the probability that user $u$ is one of the $\numVNs $  users in $\Sset$.
 
We now find expressions for the factors in \eqref{eq:ps}. %We begin with $\aS$, the expected number of ways to select $\numVNs$ VNs for $\Sset$.
We begin with $\aS$, which was derived in \cite{ivanov2015broadcast},
\begin{equation}
 \aS = \binom{m}{\numVNs}  \numVNs! \prod_{l}^{}\frac{\VNdd_l^{\numVNsL}}{\numVNsL!},
 \label{eq:aS}
\end{equation}
and stems from the multinomial distribution.
% If we consider a generic active user at some point, we do not know its' activation slot. However we know that it was activated in one of the n previous slots. Each of the n previous slots are infact equally probable for the users activation. We therefore say that there are n possibilites for the activation slot a generic user.

%If we consider a generic active user we do not know its activation slot. However we know that it was activated in one of the $n$ previous slots. Each of the $\N$ previous slots are in fact equally probable for the users activation. We therefore say that there are $\N$ possibilities for the activation slot of a generic user 
%In order to find an expression for $\bS$ we make the simplifying assumption that the selected slots of $\Sset$ are not further apart than $\N$ slots. This also means that all users of $\Sset$ are simultaneously active at some slot $i$.

Since we are considering stopping sets constrained to the slots in the range $[i,i+n-1]$ that contain $u$, the first slot of the stopping set must be $i$. According to our assumption, the remaining $\numCNs-1$ slots of $\Sset$ will be chosen with equal probability from the subsequent $\N-1$  slots. This gives, 
\begin{equation}
	\bS = \binom{\N-1}{\numCNs-1}.
	\label{eq:bS}
\end{equation}
We now consider $\dS$.  An arbitrary active user in slot $i+n-1$ has $\N$ equiprobable slots for its first replica, \IE the slots in $[i,i+n-1]$. However, the first replica of user $u$ is fixed to slot $i$. For each placement of a degree-$l$ user's first replica, there are $\binom{\N-1}{l-1}$ possible placements of its remaining replicas. Furthermore, each user places its replicas independently of other users. Thus,
\begin{equation}
	\dS = n^{-1}\prod_{l}^{}\left(\N\binom{\N-1}{l-1}\right)^{\numVNsL}.
	\label{eq:dS}
\end{equation}

An approximation of the PLR in the EF region for a given FA-CSA system is given by evaluating \eqref{eq:PLR} for some finite set of minimal stopping sets $\Astar$ which dominate the performance in the EF using \eqref{eq:ps}-\eqref{eq:dS}. 

An EF approximation can be derived in a  similar way for FS-CSA. The expressions for $\bS$ and $\dS$, however, are slightly different \cite{ivanov2015error}. We state them here for completeness,
\begin{equation}
\bSFS = \binom{\N}{\numCNs} \quad \text{and} \quad \dSFS = \prod_{l}^{}\binom{\N}{l}^{\numVNsL}.
\label{eq:bdS_FS}
\end{equation}
Evaluating \eqref{eq:PLR} and replacing $\bS$ and $\dS$ by $\bSFS$ and $\dSFS$ respectively,  yields the EF approximation for FS-CSA derived in \cite{ivanov2015broadcast} (with the addition of the expectation over the Poisson distributed RV representing the number of users per frame).
Using \eqref{eq:ps}-\eqref{eq:bdS_FS}, it is possible to rewrite \eqref{eq:PLR} after some simplifications as,  
\begin{equation}
	\PLR \approx \!\sum_{\Sset \in \Astar}^{} \!\phi(\Sset) \psi(\Sset) \numVNs \iso \binom{\N}{\numCNs}\prod_{l}^{} \frac{\VNdd_l^{\numVNsL}}{\numVNsL!} \binom{\N}{l}^{-\numVNsL} \!\!,
\end{equation}
where
\begin{equation}
	\phi (\Sset) = \sum_{k=0}^{\numVNs-1} (-1)^{\numVNs-1+k} \frac{(\numVNs-1)!}{k!} (\N\G)^k
\end{equation}
and
\begin{equation}
	\psi(\Sset) = \begin{cases}
			\numCNs \prod_{d}^{}d^{-\numVNsD} & \text{for FA} \\
			1& \text{for FS} .
		\end{cases}
\end{equation}
%%%%%%%%%%%%%%%%%%%%%%%%%%%%%%%%%%%%%%%%%%%%%%%%%%%%%%%%%%%%%%%%%%%%%%%%%%%%%%%%%%%%%%%%%%%%%%%%%%%%%%%%%%%%%%%%%%%%%%%%%%%%%
\section{Numerical Results}
\label{sec:Results}

In Table~\ref{tab:DEThreshold}, we give iterative decoding thresholds for FA-CSA  with and without boundary effect, denoted by $\gstarfaBE$ and $\gstarfanoBE$, respectively, for several regular VN-degree distributions and $\VNddOpt(x)=0.86x^3+0.14x^8$.  $\VNddOpt(x)$  was obtained in \cite{ivanov2015broadcast} for FS-CSA by a joint optimization of the EF and the threshold. 
\begin{table}[!t]
	\addtolength{\tabcolsep}{-0.7mm}
	\scriptsize
	\caption{DE Thresholds for FA-CSA and FS-CSA}
	\vspace{-3ex}
	\begin{center}\begin{tabular}{lccccccc}
			\hline
			\toprule
			$\VNdd(x)$ & $x^3$  & $x^4$ & $x^5$ & $x^6$ & $x^7$ & $x^8$ & $\VNddOpt(x)$\\
			\otoprule%\hline
			$\gstarfaBE$   & $0.917$  & $0.976$ & $0.992$ &  $0.997$  &  $0.998$  &  $0.999$    & $0.963$\\[0.5mm]
			$\gstarfanoBE$  & $0.818$ &$0.772$ & $0.701$ & $0.637$ & $0.581$ &$0.534$ & $0.851$\\[0.5mm]
			$\gstarfs$  & $0.818$ &$0.772$ &$0.701$ & $0.637$ & $0.581$ & $0.534$ & $0.851$\\[0.5mm]
			\bottomrule
		\end{tabular} \end{center}
		\label{tab:DEThreshold} % is used to refer this table in the text 
		\vspace{-4ex}
	\end{table}
We observe that if no active users are present at time $i=0$, the decoding threshold improves significantly with respect to the case where there are already active users at time $i=0$. This is due to a boundary effect (thus its name) caused by the lower degree of the CNs for $i=1,\ldots,n-1$, which results in a wave-like decoding effect similar to that of spatially coupled LDPC (SC-LDPC) codes. Furthermore, for FA-CSA with boundary effect, the decoding threshold improves with increasing VN degree, whereas the opposite occurs for the system without boundary effect. This behavior is similar to that of regular LDPC codes, where a larger VN degree improves the threshold for SC-LDPC codes but has the opposite effect for uncoupled LDPC codes.
% behavior is well known for  and LDPC codes respectively. %The lower degree CNs at the boundary will allow for a \textit{decoding wave} to emerge. This wave  moves over all nodes in the system and eventually results in the resolution of every VN. 
Spatially-coupled (frame-synchronous) CSA has been investigated in \cite{liva2012spatially}, where similar improvement of the iterative decoding threshold was observed. For FA-CSA with boundary effect, however, the \emph{spatially-coupled structure} is not enforced as in \cite{liva2012spatially}, but it is inherent to the system model.

%We observe that the system with boundary effect yields a higher decoding threshold, and that the threshold improves with increasing average VN-degree. Whereas when there is no boundary effect, the degree decreases the threshold. 

We also give in Table~\ref{tab:DEThreshold} the decoding thresholds for FS-CSA, denoted by $\gstarfs$. FA-CSA with boundary effect yields significantly better thresholds than FS-CSA. Interestingly, the thresholds for FS-CSA and FA-CSA without boundary effect are identical. Indeed, it can be shown that the two systems have identical degree distributions $\VNdd(x)$ and $\CNdd(x)$, although the node connectivity is slightly different.  

%This is  expected, since the CN-degree distributions $\CNdd(x)$ are the same for both systems. However, the systems have different connectivity and are therefore not identical.

\begin{figure}
	\centering
	\input{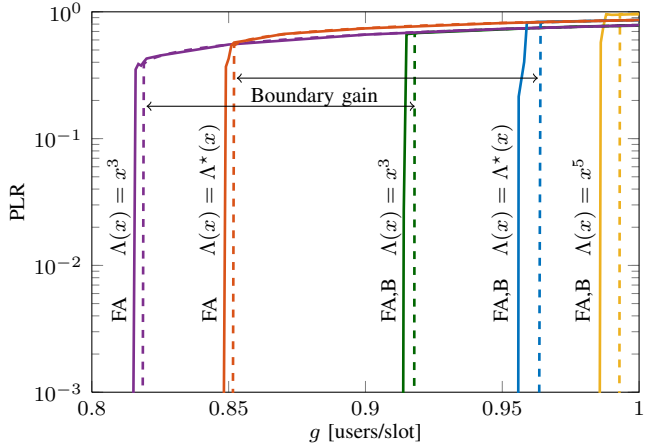}
	\vspace{-2ex}
	\caption{DE (dashed lines) and simulation results for $\N=10^5$ (solid lines) of the PLR of FA-CSA with and without boundary effect.}
	\label{fig:thresholds}
	\vspace{-2ex}
\end{figure}

In \figref{thresholds}, we plot the PLR of FA-CSA with boundary effect obtained from DE (dashed lines) together with simulation results for $\N=10^5$ (solid lines), for  $\VNdd(x)=x^k$ with $k=3$ and $5$, and $\VNdd(x)=\VNddOpt(x)$. The simulations results are in good agreement with the DE results.

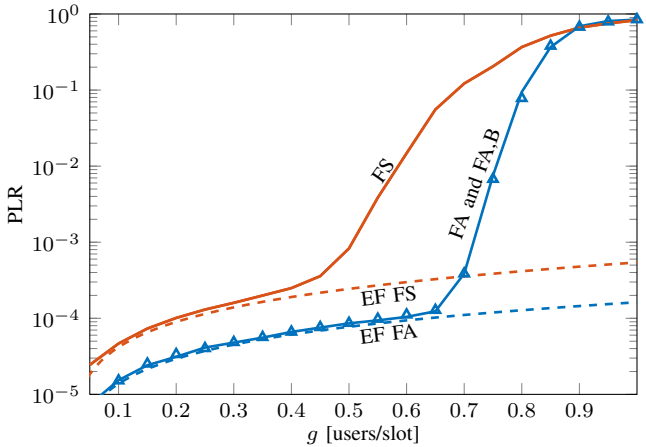
\begin{figure}[!t]
	\centering
\definecolor{mycolor1}{rgb}{0.00000,0.44700,0.74100}%
\definecolor{mycolor2}{rgb}{0.85098,0.32549,0.09804}%
\definecolor{mycolor3}{rgb}{0.00000,0.44706,0.74118}%
\begin{tikzpicture}
\pgfplotsset{every tick label/.append style={font=\footnotesize}}
\begin{axis}[%
width=\columnwidth,
height=0.75*\columnwidth,
at={(1.387in,0.821in)},
xmin=0.05,
xmax=1,
xlabel={\footnotesize $g$ [users/slot]},
xlabel style={
	yshift=1.5ex,
	name=label},
xtick ={0,0.1,0.2,0.3,0.4,0.5,0.6,0.7,0.8,0.9},
ymode=log,
ymin=1e-05,
ymax=1,
yminorticks=true,
ylabel={\footnotesize PLR},
ylabel style={
	yshift=-1.5ex,
	name=label},
axis background/.style={fill=white},
legend style={at={(0,1)},anchor=north west,legend cell align=left,align=left,draw=white!15!black}
]
\addplot [color=mycolor1,solid,line width=1pt,forget plot] % With the BOUNDARY EFFECT!
table[row sep=crcr]{%
	0.05	7.35e-06\\
	0.1	1.537e-05\\
	0.15	2.416e-05\\
	0.2	3.12e-05\\
	0.25	4.123e-05\\
	0.3	4.791e-05\\
	0.35	5.568e-05\\
	0.4	6.618e-05\\
	0.45	7.549e-05\\
	0.5	8.58799991412e-05\\
	0.55	9.389e-05\\
	0.6	0.000104417910298771\\
	0.65	0.000123482129876084\\
	0.7	0.000386249576863589\\
	0.75	0.00752572292094379\\
	0.8	0.0944831300371319\\
	0.85	0.366901460121799\\
	0.9	0.689607613268051\\
	0.95	0.806321561038542\\
	1	0.84573748308525\\
};
%\addlegendentry{\footnotesize ,B};

\addplot [color=mycolor1,only marks,line width=1pt,mark=triangle,mark options={solid,fill=gray},forget plot] % WITHOUT BOUNDARY EFFECT
table[row sep=crcr]{%
	0.05	7.42e-06\\
	0.1	1.516e-05\\
	0.15	2.521e-05\\
	0.2	3.363e-05\\
	0.25	4.03099995969e-05\\
	0.3	4.845e-05\\
	0.35	5.645e-05\\
	0.4	6.635e-05\\
	0.45	7.624e-05\\
	0.5	8.64899991351e-05\\
	0.55	9.798e-05\\
	0.6	0.00011149\\
	0.65	0.0001281\\
	0.7	0.00038643\\
	0.75	0.00675677319198666\\
	0.8	0.0780965679682241\\
	0.85	0.380361190986961\\
	0.9	0.675315203371173\\
	0.95	0.793167721569187\\
	1	0.847637211273575\\
};

%\addlegendentry{\footnotesize FA};
\addplot [color=mycolor2,solid,line width=1pt]
table[row sep=crcr]{%
	0.05	2.44e-05\\
	0.1	4.66099986017e-05\\
	0.15	7.33299992667e-05\\
	0.2	0.000101059514171215\\
	0.25	0.000130268364423437\\
	0.3	0.000159814630370974\\
	0.35	0.000199280680631697\\
	0.4	0.000249410433984879\\
	0.45	0.000358119051306821\\
	0.5	0.000828206361225745\\
	0.55	0.00385343882953715\\
	0.6	0.014761502588771\\
	0.65	0.055534925482328\\
	0.7	0.121839247977847\\
	0.75	0.204595588235294\\
	0.8	0.367311843693974\\
	0.85	0.518419290020094\\
	0.9	0.661092824226465\\
	0.95	0.751163838414176\\
	1	0.829302517540239\\
};
%\addlegendentry{\footnotesize FS};

\addplot [color=mycolor2,solid,line width=1pt]
table[row sep=crcr]{%
	0.05	2.44e-05\\
	0.1	4.66099986017e-05\\
	0.15	7.33299992667e-05\\
	0.2	0.000101059514171215\\
	0.25	0.000130268364423437\\
	0.3	0.000159814630370974\\
	0.35	0.000199280680631697\\
	0.4	0.000249410433984879\\
	0.45	0.000358119051306821\\
	0.5	0.000828206361225745\\
	0.55	0.00385343882953715\\
	0.6	0.014761502588771\\
	0.65	0.055534925482328\\
	0.7	0.121839247977847\\
	0.75	0.204595588235294\\
	0.8	0.367311843693974\\
	0.85	0.518419290020094\\
	0.9	0.661092824226465\\
	0.95	0.751163838414176\\
	1	0.829302517540239\\
};

\addplot [color=mycolor3,dashed,line width=1pt]
table[row sep=crcr]{%
0.01	5.61158919426908e-07\\
0.02	1.73278804159215e-06\\
0.03	3.13033003657333e-06\\
0.04	4.61231896377527e-06\\
0.05	6.12671720706205e-06\\
0.06	7.65438599232475e-06\\
0.07	9.1882893530745e-06\\
0.08	1.07258436969263e-05\\
0.09	1.22661033683924e-05\\
0.1	1.38087252752955e-05\\
0.11	1.53535879961122e-05\\
0.12	1.69006516573952e-05\\
0.13	1.84499063855582e-05\\
0.14	2.00013533433468e-05\\
0.15	2.15549977535462e-05\\
0.16	2.31108463325467e-05\\
0.17	2.46689063462054e-05\\
0.18	2.62291852625166e-05\\
0.19	2.77916906238371e-05\\
0.2	2.93564299998798e-05\\
0.21	3.09234109704215e-05\\
0.22	3.24926411189414e-05\\
0.23	3.40641280302806e-05\\
0.24	3.56378792897811e-05\\
0.25	3.72139024829695e-05\\
0.26	3.87922051954401e-05\\
0.27	4.0372795012812e-05\\
0.28	4.19556795207137e-05\\
0.29	4.35408663047769e-05\\
0.3	4.51283629506347e-05\\
0.31	4.67181770439206e-05\\
0.32	4.83103161702681e-05\\
0.33	4.9904787915311e-05\\
0.34	5.1501599864683e-05\\
0.35	5.31007596040179e-05\\
0.36	5.47022747189494e-05\\
0.37	5.63061527951112e-05\\
0.38	5.79124014181371e-05\\
0.39	5.95210281736608e-05\\
0.4	6.1132040647316e-05\\
0.41	6.27454464247366e-05\\
0.42	6.43612530915562e-05\\
0.43	6.59794682334086e-05\\
0.44	6.76000994359275e-05\\
0.45	6.92231542847467e-05\\
0.46	7.08486403654999e-05\\
0.47	7.24765652638209e-05\\
0.48	7.41069365653434e-05\\
0.49	7.57397618557011e-05\\
0.5	7.73750487205278e-05\\
0.51	7.90128047454572e-05\\
0.52	8.06530375161231e-05\\
0.53	8.22957546181591e-05\\
0.54	8.39409636371992e-05\\
0.55	8.55886721588769e-05\\
0.56	8.7238887768826e-05\\
0.57	8.88916180526803e-05\\
0.58	9.05468705960736e-05\\
0.59	9.22046529846395e-05\\
0.6	9.38649728040118e-05\\
0.61	9.55278376398242e-05\\
0.62	9.71932550777106e-05\\
0.63	9.88612327033045e-05\\
0.64	0.00010053177810224\\
0.65	0.00010220489886015\\
0.66	0.00010388060256267\\
0.67	0.000105558896795431\\
0.68	0.00010723978914407\\
0.69	0.000108923287194218\\
0.7	0.00011060939853151\\
0.71	0.00011229813074158\\
0.72	0.000113989491410061\\
0.73	0.000115683488122587\\
0.74	0.000117380128464791\\
0.75	0.000119079420022309\\
0.76	0.000120781370380772\\
0.77	0.000122485987125816\\
0.78	0.000124193277843073\\
0.79	0.000125903250118178\\
0.8	0.000127615911536764\\
0.81	0.000129331269684465\\
0.82	0.000131049332146915\\
0.83	0.000132770106509747\\
0.84	0.000134493600358596\\
0.85	0.000136219821279094\\
0.86	0.000137948776856877\\
0.87	0.000139680474677576\\
0.88	0.000141414922326827\\
0.89	0.000143152127390262\\
0.9	0.000144892097453517\\
0.91	0.000146634840102223\\
0.92	0.000148380362922016\\
0.93	0.000150128673498529\\
0.94	0.000151879779417395\\
0.95	0.000153633688264249\\
0.96	0.000155390407624724\\
0.97	0.000157149945084454\\
0.98	0.000158912308229072\\
0.99	0.000160677504644212\\
1	0.000162445541915509\\
};

\addplot [color=mycolor2,dashed,line width=1pt]
table[row sep=crcr]{%
0.01	1.68451941035384e-06\\
0.02	5.20521206457939e-06\\
0.03	9.41046748141335e-06\\
0.04	1.3876709520129e-05\\
0.05	1.844819090635e-05\\
0.06	2.30676930993824e-05\\
0.07	2.77142442540187e-05\\
0.08	3.23802069391313e-05\\
0.09	3.70628491803925e-05\\
0.1	4.17612436201812e-05\\
0.11	4.64751267822616e-05\\
0.12	5.12044794187108e-05\\
0.13	5.59493721281814e-05\\
0.14	6.0709908562034e-05\\
0.15	6.54862045310409e-05\\
0.16	7.02783803191717e-05\\
0.17	7.50865578559937e-05\\
0.18	7.99108596764557e-05\\
0.19	8.47514085382138e-05\\
0.2	8.96083272808535e-05\\
0.21	9.44817387741006e-05\\
0.22	9.93717658987688e-05\\
0.23	0.000104278531539751\\
0.24	0.00010920215858344\\
0.25	0.000114142769916781\\
0.26	0.000119100488426924\\
0.27	0.00012407543700109\\
0.28	0.000129067738526531\\
0.29	0.000134077515890508\\
0.3	0.000139104891980284\\
0.31	0.000144149989683126\\
0.32	0.0001492129318863\\
0.33	0.000154293841477072\\
0.34	0.00015939284134271\\
0.35	0.000164510054370479\\
0.36	0.000169645603447646\\
0.37	0.000174799611461479\\
0.38	0.000179972201299242\\
0.39	0.000185163495848204\\
0.4	0.000190373617995631\\
0.41	0.000195602690628789\\
0.42	0.000200850836634946\\
0.43	0.000206118178901367\\
0.44	0.000211404840315319\\
0.45	0.000216710943764068\\
0.46	0.000222036612134883\\
0.47	0.000227381968315028\\
0.48	0.000232747135191772\\
0.49	0.000238132235652379\\
0.5	0.000243537392584118\\
0.51	0.000248962728874254\\
0.52	0.000254408367410054\\
0.53	0.000259874431078785\\
0.54	0.000265361042767714\\
0.55	0.000270868325364106\\
0.56	0.00027639640175523\\
0.57	0.00028194539482835\\
0.58	0.000287515427470735\\
0.59	0.00029310662256965\\
0.6	0.000298719103012362\\
0.61	0.000304352991686138\\
0.62	0.000310008411478244\\
0.63	0.000315685485275948\\
0.64	0.000321384335966515\\
0.65	0.000327105086437212\\
0.66	0.000332847859575306\\
0.67	0.000338612778268064\\
0.68	0.000344399965402751\\
0.69	0.000350209543866636\\
0.7	0.000356041636546984\\
0.71	0.000361896366331061\\
0.72	0.000367773856106136\\
0.73	0.000373674228759473\\
0.74	0.000379597607178341\\
0.75	0.000385544114250004\\
0.76	0.000391513872861731\\
0.77	0.000397507005900788\\
0.78	0.000403523636254441\\
0.79	0.000409563886809956\\
0.8	0.000415627880454602\\
0.81	0.000421715740075643\\
0.82	0.000427827588560347\\
0.83	0.000433963548795981\\
0.84	0.00044012374366981\\
0.85	0.000446308296069103\\
0.86	0.000452517328881124\\
0.87	0.000458750964993142\\
0.88	0.000465009327292422\\
0.89	0.000471292538666231\\
0.9	0.000477600722001835\\
0.91	0.000483934000186503\\
0.92	0.000490292496107499\\
0.93	0.00049667633265209\\
0.94	0.000503085632707544\\
0.95	0.000509520519161127\\
0.96	0.000515981114900105\\
0.97	0.000522467542811745\\
0.98	0.000528979925783314\\
0.99	0.000535518386702078\\
1	0.000542083048455304\\
};

\node[above,left,rotate=55] at (axis cs:0.58,0.015 ){\footnotesize FS};
\node[above,left,rotate=70] at (axis cs:0.76,0.04 ){\footnotesize FA and FA,B};
\end{axis}
%\node(fs) at (7.8,5.3) {\footnotesize FS};
%\node(fa) at (8.8,5.3) {\footnotesize FA-$\BE$};
%\draw[black, -] (fa.east) -- (9.7,5.3); 
%\node(fanobe) at (8.8,4.93){\footnotesize FA-$\noBE$};
%\draw[black, -] (fanobe.east) -- (9.4,4.93); 
\node[rotate=7](ef1) at (7.5,2.9) {\footnotesize EF FA};
\node[rotate=7] (ef2)at (7.5,3.4) {\footnotesize EF FS};
\end{tikzpicture}%
	\vspace{-2ex}
	\caption{PLR performance of FS-CSA and FA-CSA for $\VNddOpt(x)$ and $\N=100$. Blue markers show results for FA-CSA without boundary effect and the solid blue line shows results for FA-CSA with boundary effect.}
	\label{fig:plr}
	\vspace{-2ex}
\end{figure}

In \figref{plr}, we plot the simulated PLR as a function of $g$ for FA-CSA with and without boundary effect and FS-CSA, for $\VNddOpt(x)$ and  $\N=100$. The EF predictions as derived in Section~\ref{sec:EF} are also shown. We observe that FA-CSA outperforms FS-CSA in both the EF and  WF regions. Surprisingly, the PLR curves for FA-CSA with and without boundary effect are on top of each other. This implies that for short block lengths the boundary effect does not translate into better performance. However, with increasing $n$ the two curves move apart from each other, in agreement with the DE prediction.  % Nevertheless the performance for FA-CSA in the EF is better by a factor of around $3$, in addition to this, t waterfall performance is also better for FA-CSA. 

\begin{figure}[!t]
	\centering
	\input{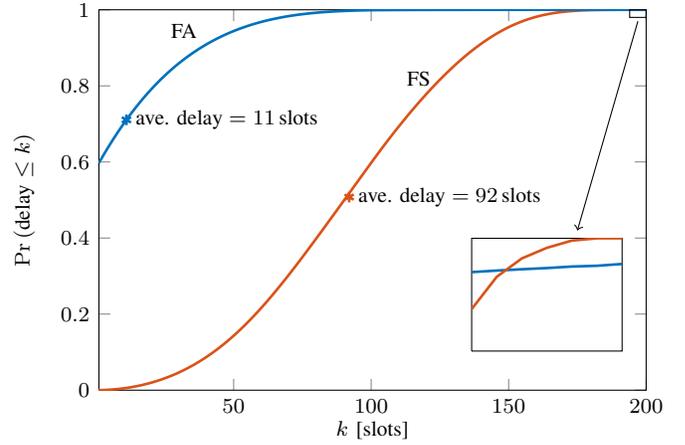}
	\vspace{-4ex}
	\caption{CDF of the delay of received packets in terms number of slots for FS-CSA and FA-CSA, for $\VNddOpt(x)$, $\N=100$ and,  $\G=0.4$.}
	\label{fig:delay_cdf}
	\vspace{-2ex}
\end{figure}

Finally, in \figref{delay_cdf} we depict the cumulative distribution function (CDF) for the delay of resolved packets for FA-CSA without boundary effect and FS-CSA, with parameters  $\N=100$, $\G=0.4$, and $\VNdd(x)=\VNddOpt(x)$. Here the delay is defined as the number of slots between a user joining the system and the successful decoding of its packet. We observe that FA-CSA provides a lower average delay than FS-CSA. This is true in general, \IE for any $\N$, $\G$, and $\VNdd(x)$. However, we remark that the maximum delay of FA-CSA is $\N+\NRX$, which can be greater than $2\N-1$, the maximum delay for FS-CSA. This explains the crossing in \figref{delay_cdf}. Nevertheless, it is important to point out that the probability of a delay greater than $2\N-1$ for FA-CSA is very small.

\balance
\section{Conclusions}
We analyzed the asymptotic and finite frame length performance of frame-asynchronous CSA. We derived the DE that characterizes the system performance as the frame length tends to infinity, and showed that if a boundary effect is present, the threshold for FA-CSA is greatly improved. Furthermore, we derived analytical approximations of the EF in the finite frame length regime. We showed that FA-CSA outperforms FS-CSA in terms of  asymptotic threshold, as well as EF, waterfall performance, and average delay for finite frame length.

\bibliographystyle{IEEEtran}
\bibliography{library,IEEEabrv,IEEEfull}
\end{document}